\documentclass[9pt,sigconf,letterpaper,nonacm]{acmart}

\usepackage[utf8]{inputenc}
\usepackage[T1]{fontenc}
\usepackage{amsmath,amssymb,bm}
\usepackage{multirow}
\usepackage{listing}

\usepackage{tikz}
\usetikzlibrary{bayesnet}

\DeclareMathOperator*{\argmax}{arg\,max}

\def\Pprob{\mathbb{P}}
\def\un{{\mbox{\rm 1\hspace{-0.25em}I}}}

\begin{document}
\title[Large-Scale Characterization and Segmentation of Internet Path Delays with Infinite HMMs]{Large-Scale Characterization and Segmentation of Internet Path Delays with Infinite HMMs}

\author[M. Mouchet]{Maxime Mouchet}
\affiliation{
  \institution{IMT Atlantique, Lab-STICC}
}
\email{maxime.mouchet@imt-atlantique.fr}

\author[S. Vaton]{Sandrine Vaton}
\affiliation{
  \institution{IMT Atlantique, Lab-STICC}
}
\email{sandrine.vaton@imt-atlantique.fr}

\author[T. Chonavel]{Thierry Chonavel}
\affiliation{
  \institution{IMT Atlantique, Lab-STICC}
}
\email{thierry.chonavel@imt-atlantique.fr}

\author[E. Aben]{Emile Aben}
\affiliation{
  \institution{RIPE NCC}
}
\email{emile.aben@ripe.net}

\author[J. den Hertog]{Jasper den Hertog}
\affiliation{
  \institution{RIPE NCC}
}
\email{jdenhertog@ripe.net}

\begin{abstract}
Round-Trip Times are one of the most commonly collected performance metrics in computer networks.
Measurement platforms such as RIPE Atlas provide researchers and network operators with an unprecedented amount of historical Internet delay measurements.
It would be very useful to automate the processing of these measurements (statistical characterization of paths performance, change detection, recognition of recurring patterns, etc.).
Humans are pretty good at finding patterns in network measurements but it can be difficult to automate this to enable many time series being processed at the same time.
In this article we introduce a new model, the HDP-HMM or infinite hidden Markov model, whose performance in trace segmentation is very close to human cognition. This is obtained at the cost of a greater complexity and the ambition of this article is to make the theory accessible to network monitoring and management researchers. We demonstrate that this model provides very accurate results on a labeled dataset and on RIPE Atlas and CAIDA MANIC data. This method has been implemented in Atlas and we introduce the publicly accessible Web API.

\end{abstract}


\maketitle

\section{Introduction}

\subsection{Scope of the paper}

Network management has traditionally been entrusted to humans. But this mode of operation is expensive, error-prone, and slow to adapt to changes. The task of human experts is very complex because of the large number and heterogeneity of equipments, as well as the wide variety of applications.

We believe that the future of network management is in automation, or driverless (autonomous) networks. \cite{boutaba_comprehensive_2018,Google-zerotouch,Juniper-selfdriving,ONAP}.
For self-driving networks to become reality, it is necessary to rely on recent machine learning techniques to extract information from network measurements and automate decision-making.
Different needs should be addressed: statistical characterization, prediction, detection of changes or anomalies, classification, etc. The results should be reliable and accurate to automate decision-making related to network management or to security and resilience and the analysis should be scalable and fully automated (no human intervention).

\begin{figure}[h]
    \centering
    \includegraphics[width=\linewidth]{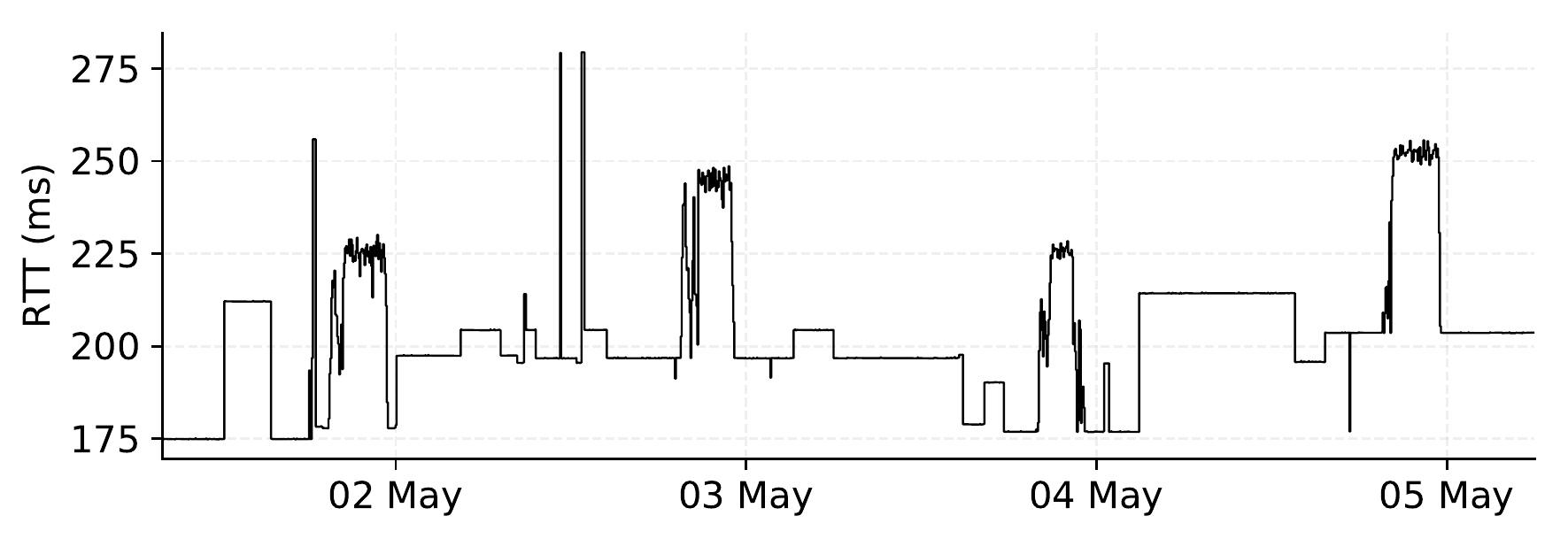}
    \caption{RTT between two RIPE Atlas anchors from May 1st to May 5th, 2018.}
    \label{fig:sample-raw-rtt}
\end{figure}

Delay is an important performance metric. In particular it is easy to measure Round Trip Time (RTT) and there is a good availability of data from measurement infrastructures at the Internet scale like RIPE Atlas \cite{staff_ripe_2015}. Humans are pretty good at finding patterns in this latency data (try it for yourself in Figure \ref{fig:sample-raw-rtt}), but it is difficult to automate this which would allow many time series being processed at the same time.

In this article we propose to use a Hierarchical Dirichlet Process Hidden Markov Model (HDP-HMM), also called infinite HMM, or nonparametric Bayesian HMM. 

This model mimics human cognition very well (in terms of segmentation of the series, recognition of different states, etc.).

These models are used for the segmentation of audio sequences for which they give very good results for speaker recognition \cite{fox2011sticky}. These recent techniques are more complex than standard HMMs but they are worth the effort.

The goal of the article is to recall the major principles of infinite HMMs and apply this theory to network measurement data. Whereas \cite{Neal00markovchain,teh_hierarchical_2006,fox2011sticky} are written for statisticians, we want to make the
theory accessible to a wider audience and show the potential of this model for automating the analysis of a wide variety of delay time series. The method has been implemented in RIPE Atlas to automate the processing of anchor to anchor RTT measurements, and a Web API is available. The article introduces the API and an accompanying notebook is provided to help taking control of the API\footnote{\url{https://github.com/maxmouchet/atlas-trends-demo}.}.

\subsection{State of the art}

Network delay modeling and prediction is a well-studied problem. Some of the simplest models assume independent observations and can be used to detect anomalous delay values. They, however, cannot predict the delay or find recurring patterns in a delay series since they do not account for time dependencies.
Such models include the Pareto distribution \cite{Zhang2007}, mixtures of Weibull \cite{Hernandez2006} or Normal distributions \cite{Sato2005}.

More complex time series models have been used for short-term delay predictions (from seconds to minutes), with applications such as telerobotics. These include autoregressive models \cite {Yang2005, Kamrani2005} and deep neural networks \cite {Belhaj2009, dong_round_2019}. As a drawback their parameters are more difficult to interpret and they do not provide a segmentation of the data.

HMMs are another kind of time-series model that can model different delay distributions and the dynamics between them.
In \cite{Salamatian2001} a discrete-time HMM is used to model packet losses, while in \cite{WEI2002129} a continuous-time HMM is used to model both packet losses and delays. In \cite{DAINOTTI20082645} a HMM is used to model inter-packet times and packet sizes. HMMs have few parameters and those are easily interpretable (state transition probabilities, means, variances, ...).

However, standard HMMs require to use heuristics to determine the number of hidden states. To remedy this problem we use a nonparametric HMM for which the number of hidden states is inferred from the data.
A nonparametric mixture model has been used in the past to model the delay of a set of hosts measured over disjoint time intervals \cite {Fontugne2015}. In comparison our model is a nonparametric HMM and concerns the delay between two hosts over a large and continuous time interval, from a few hours to a few weeks.

We first introduced the use of the HDP-HMM for RTT time series in \cite{8717870}. This article expands on the statistical theory behind the model, describes two new applications (CAIDA MANIC measurements, and anomaly detection), and introduces a RIPE Atlas API for time series segmentation as a service.

\subsection{Structure of the article}

The paper is structured as follows. Section \ref{sec:markov-model} is a reminder on mixture models (MM) and hidden Markov models (HMM). In section \ref{sec:npbayesian} we describe their nonparametric Bayesian counterparts, the Dirichlet Process MM (DPMM) and the Hierarchical Dirichlet Process HMM (HDP-HMM, or infinite HMM). In the same section, we explain how to automatically calibrate these models, that is how their parameters can be inferred from measurements without human intervention. 

In section \ref{sec:first-look} the accuracy of the model is demonstrated on a dataset that has been labeled by humans, as well as on some RIPE Atlas RTT time series where we discuss the matching between routing configurations (from traceroutes) and states learned by the statistical model. We also briefly address the analysis of some CAIDA MANIC measurements. In Section \ref{sec:large-scale} we introduce a Web API that permits requesting the HDP-HMM analysis of anchor to anchor RTT measurements in RIPE Atlas. We also demonstrate that analyzing the frequency of state changes in RTT time series over Atlas allows a very precise detection of the moment of occurrence of events affecting large infrastructures of the Internet (such as IXPs). In Section \ref{sec:conclusion} we conclude and present our vision of the research axes to be developed in the future.

Readers that are less interested in the description of the Bayesian nonparametric context can skip most of sections \ref{sec:markov-model} and \ref{sec:npbayesian}, just reading their summaries (subsections \ref{sec:markov-model-summary} and \ref{sec:npbayesian-summary}).

\section{A reminder on mixture models and hidden Markov models}
\label{sec:markov-model}

In the next two sections our goal is to let the reader understand the HDP-HMM model, starting from simpler and more popular models such as mixtures or HMMs.

\begin{table}[h]
    \centering
    \caption{Taxonomy of models}
    \label{tab:taxonomy}
    \begin{tabular}{lccc}
    \toprule
    \textbf{Model} &  \textbf{Number} & \textbf{Time} \\
                   & \textbf{of states} & \textbf{dependency} \\
    \midrule
        Mixture Model & Fixed & No
        &\\
        Hidden Markov Model & Fixed & Yes
        &\\
        \midrule
        Dirichlet Process Mixture & Infinite & No
        &\\
        Hierarchical Dirichlet Process & &\\
        Hidden Markov Model & Infinite & Yes
        & \\
        \bottomrule
    \end{tabular}
\end{table}

\subsection{A taxonomy of statistical models}

We will start by a taxonomy of the different models discussed in the article.
Our taxonomy takes into account three criteria: (i) whether there is naturally a notion of "hidden state" in the statistical model (ii) whether time dependency is taken into account, and (iii) whether the number of states is supposed to be known (and finite) or unknown (and potentially infinite).

The RTT is stable over long periods of time (usually a few hours), and its distribution switches from one probability law to another (see Fig. \ref{fig:sample-raw-rtt}). This can be explained by IP-level routing changes, congestion, and traffic engineering at lower layers than layer 3 \cite{Pucha2007}. Propagation delays give a lower bound on the RTTs, and as routers queue lengths increase with the traffic, so do the observed RTTs. 
From a statistical point of view, it is natural to think of models with "hidden states" such as MMs or HMMs.

Knowing that the delay is stable over several hours means that, if the path quality is measured at a frequency of one "ping" every few minutes, the delay distribution remains stable for tens or hundreds of time slots. In order to have a model that can be used for prediction, it is necessary to account for this temporal dependence. This is made possible by HMMs, while mixture MMs assume independent observations.

But a classical problem in statistics with MMs or HMMs is that the order of the model is assumed to be known (and finite). In practice this hypothesis is unrealistic, in most applications considered. This is where models with Dirichlet processes (DP) priors on the number of components of the mixture, or of the HMM, find their interest. 

In the Dirichlet Process MM (DPMM) and the Hierarchical Dirichlet Process HMM (HDP-HMM), the number of model states is "infinite". And the order of the model can be learned from the measured data, as it is the case for the other parameters of the model. This is an important property to have a model that is flexible enough to adapt, without manual human intervention (initialization of algorithms, etc.), to a large number of time series.

In Table \ref{tab:taxonomy} we have summarized which properties are satisfied by which models. This justifies the choice of the HDP-HMM to characterize RTTs and to automate their processing.

This flexibility is obtained at the cost of a greater complexity of the model of inference algorithms for parameter estimation. However, we could provide an efficient implementation for it embedded in an operational API (see Section \ref{sec:Atlas-API}).

\subsection{Mixture Models}

Some of the simplest statistical models that include hidden states are mixture models.
MMs are a kind of generative statistical models used to describe data produced by different system states. For instance, in a Gaussian Mixture Model (GMM), observations $y_{1:T} = (y_1, y_2, \ldots, y_T)$ are assumed to be independent and a normal distribution is associated to each hidden state. For continuously distributed observations, conditionally to the underlying state $z_t = k \in \{1,2,\ldots,K\}$, where $K$ denotes the number of states of the model, the observation $y_t$ follows a distribution with probability density function $p_{\theta_{k}}$, where $\theta_{k}$ is a parameter vector. For example, in a GMM, $\theta_{k}$ consists of mean and variance parameters, so $\theta_{k} = (\mu_{k}, \sigma^2_{k})$ and $p_{\theta_k}(y)=\mathcal{N}(y;\mu_{k}, \sigma^2_{k})=(2\pi\sigma_{k}^2)^{-1/2}\exp\left(-\frac{(y-\mu_{k})^2}{2\sigma_{k}^2}\right)$. Finally, the data distribution writes
$p(y_t)=\sum_{k=1:K}\pi_k p_{\theta_{k}}(y_t)$
where $\pi_k$ denotes the probability that the state of an observation is $k$, that is, $\pi_k = \Pprob(z_t = k)$. 

MM parameters ${\bm \phi}=\{\pi_k,\theta_k\}_{k=1:K}$ can be estimated from measurements $y_{1:T}$ according to different criteria and algorithms. A common choice is the Maximum Likelihood Estimator (MLE) which supplies the parameters that maximize the likelihood of the observations: 
${\bm \phi}_{MLE} = \argmax_{\bm \phi} p(y_1, y_2, \ldots, y_T; \bm \phi)$.
 In general, direct maximization of the likelihood $p(y_{1:T}; \bm \phi)$ with respect to $\bm \phi$
is infeasible. The Expectation-Maximization (EM) algorithm \cite{dempster77} is a popular iterative two-steps algorithm to compute the MLE for models with incomplete data, in particular mixture models. 

\subsection{Hidden Markov Models}

Because of the independent observations assumption, the predictive ability of MMs is limited. Knowing model parameters and which state value $z_t$ has generated the last observation $y_t$ does not bring any information about the next state $z_{t+1}$. HMMs are a generalization of MMs that take into account temporal dependencies among states. These temporal dependencies are expressed through a Markov property assumed for the states that writes $p(z_{t+1}\vert z_{1:t})=p(z_{t+1}\vert z_{t})$.
Thus, the probability distribution of the next hidden state $z_{t+1}$ depends on the current state $z_t$ only. 

More formally the transition probabilities between successive states are defined via a $K\times K$ matrix ${\bf \Pi}$ with entries ${\bf \Pi}_{ij}=P(z_{t+1} = j | z_{t} = i)$. The model parameters are now ${\bm \phi}=\{{\bf \Pi},\{\theta_k\}_{k=1:K} \}$, the steady state probability vector ${\bm \pi} = (\pi_1, \ldots, \pi_K)$ being related to ${\bf \Pi}$ through the linear system ${\bm  \pi} {\bf \Pi} ={\bm \pi}$ and ${\bm \pi e}=1$, where ${\bm e}= (1,\ldots,1)^T$.

The MLE of HMM parameters can be estimated using a variant of the EM algorithm known as the Baum-Welch (or \textit{forward-backward}) algorithm \cite{hmm-rabiner}. While easy to implement and well-studied, this approach is prone to overfitting on noisy data or data with few samples. Furthermore this method requires the number $K$ of hidden states to be known, which is usually not the case when studying RTTs. 

\subsection{Limitations of vanilla MMs and HMMs}
\label{sec:model-selection}
Classical mixtures and HMMs are parametric models, meaning that they have a set of parameters with fixed size. This is a major difficulty when estimating HMM parameters as often the number of hidden states is not a priori known.

One could estimate models for different numbers of states, but the maximum of the likelihood would increase with the number of states as a model of order $K$ is a degenerated case of model of order $K+1$. A classical approach consists in penalizing the MLE optimization criterion by adding a penalty term to the log-likekihood such as the AIC \cite{Akaike1974} or the BIC \cite{schwarz1978} criterions and by selecting the model that minimizes this penalized criterion. Another approach is to use nonparametric models with unbounded number of parameters.

Another limitation of parametric models is that the EM algorithm usually used to tune the parameters of the model is sensitive to the choice of its initialization point. Appropriate initialization strategies must be used otherwise it may converge to a local but non-global maximum of the likelihood.

Because of these limitations standard MMs or HMMs cannot be used on a large scale to analyze Internet measurements. In what follows we introduce a new approach for RTT measurement analysis, based on nonparametric Bayesian models, and more particularly the HDP-HMM.

\subsection{Section summary}
\label{sec:markov-model-summary}

MMs and HMMs are interesting for characterizing time series of RTTs. They are designed to model phenomena that change state from time to time and in which the value of the observations, here the RTTs, noisily depends on the hidden states. One can imagine that different hidden states result from different routing configurations, traffic engineering choices, or link loads. However, these models are too simple to characterize a large variety of RTT series and not suitable for automating their processing at large scale.

We propose to use a more generic model, the HDP-HMM. This model does not make assumptions about the number of states of the system, contrary to vanilla mixtures or HMMs, and it is possible to learn the number of states from the data itself. Contrary to DPMMs it also takes into account time dependency and makes it possible to account for the RTT distribution being stable for a long period of time.

\section{Nonparametric Bayesian approach}
\label{sec:npbayesian}

A more formal approach to models with an unknown number of components can be found in Bayesian statistics. The Bayesian framework allows one to specify models with several layers of uncertainty and to perform inference of the parameters in a systematic way. We will make better use of this flexibility to estimate HMMs from RTT series where neither the number of states, nor the probability distribution in each state is known.

\subsection{Bayesian setting}

In the MLE approach, estimates of the parameters are inferred from data. In contrast, Bayesian approaches make use of prior distributions upon the model parameters, and output a posterior probability distribution over the model parameters. These prior distributions can account for prior knowledge upon the parameters distributions.

When the dimension of the model is unknown as for MMs or HMMs with unknown order $K$, one can resort to nonparametric Bayesian approaches, where the number of components of the model is inferred from the data itself. 

Bayesian inference can be performed from the posterior likelihood which is defined as 
$ p({\bm \phi}\vert y_{1:T})\propto p(y_{1:T}\vert{\bm \phi})p({\bm \phi})$
where $p(y_{1:T}\vert{\bm \phi})$ is the likelihood of the data $y_{1:T}$, $p({\bm \phi})$ is a \textit{prior distribution} and $\propto$ denotes proportionality.

In general, a direct maximization of the posterior likelihood $p({\bm \phi}\vert y_{1:T})$ with respect to $\bm \phi$ is not feasible as $p({\bm \phi}\vert y_{1:T})$ can be quite complex. Note, however, that there are situations where the likelihood and the prior distribution are such that posterior distribution belongs to the same family as the prior. In this case, the prior is said to be conjugate. Using conjugate priors, when possible, often makes inference simpler.

Markov Chain Monte Carlo (MCMC) techniques, and in particular Gibbs sampling, can be used in very general situations for inference \cite{Robert98montecarlo}. Alternatively, variational Bayesian methods can be considered (\cite{MacKay_2002}, chap. 33). The principle of MCMC methods is to use simulations to draw a large number of samples $\bm \phi$ from the posterior distribution $p(\bm \phi \vert y_{1:T})$. 

\subsection{Dirichlet Processes and DP mixtures}

Modelling a HMM with an infinite number of states is generally achieved by means of a Dirichlet process (DP) prior. DPs have been introduced by Ferguson \cite{Ferguson73} in 1973 and have first been applied to mixture models with an unknown number of components in \cite{Antoniak74}. The extension to the modelling of HMMs has first been defined in 2002 in \cite{beal_infinite_2002}. More recently this has been formalized in the framework of hierarchical Dirichlet processes (HDP) in \cite{teh_hierarchical_2006} where HDP-HMMs have been introduced. These models are called nonparametric Bayesian, meaning that they are Bayesian and involve parameter spaces of infinite dimension \cite{GERSHMAN2012}.

A Dirichlet Process (DP) is a stochastic process $G\sim DP(\alpha,H)$, the realizations of which are probability distributions. It is parameterized by a concentration parameter $\alpha$ and a base distribution $H$.
It can be seen as a process indexed by partitions $(A_1,\ldots,A_n)$ ($n>0$) of the space $E$ on which $H$ is defined, with $n$-variate Dirichlet random realizations:
\begin{equation}
    (G(A_1),\ldots,G(A_n))\sim {\text{\bf Dir}}(\alpha H(A_1),\ldots,\alpha H(A_n)).
    \label{eq:DP-basedefinition}
\end{equation} 
Here ${\text{\bf Dir}}(\alpha_1,\ldots,\alpha_n)$ denotes the $n$-variate Dirichlet \textit{distribution} with parameters $\alpha_{1:n}=(\alpha_1,\ldots,\alpha_n)$, that is to say the probability distribution with density function: 
\begin{equation}
    p(x_{1:n}; \alpha_{1:n}) = \frac{1}{B(\bm \alpha)} \un_{\{1\}}(\sum_{i=1:n}x_i)\prod_{k=1:n} x_i^{\alpha_i-1}\un_{[0,1]}(x_i)
\end{equation}
where $\un_A(x)=1$ if $x\in A$ and 0 otherwise, 
and $B(\bm \alpha)$ is a normalization factor.

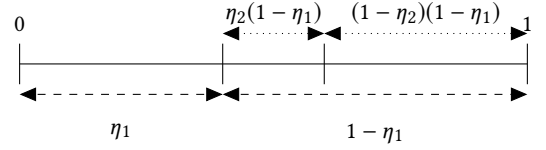
\begin{figure}
\centering
\begin{tikzpicture}[scale=1.35]
  \draw (0,0) -- (5,0) ;
  \node at (0,0.4) {$0$};
  \node at (5,0.4) {$1$};
  \draw (0,-0.2) -- (0,0.2);
  \draw (5,-0.2) -- (5,0.2);
  \draw (2,-0.2) -- (2,0.2);
  \draw (3,-0.2) -- (3,0.2);
  \draw [<->, dashed] (0,-0.3) -- (2,-0.3);
  \draw (1,-0.5) node[below]{$\eta_1$} ;
  \draw [<->, dashed] (2,-0.3) -- (5,-0.3);
  \draw (3.5,-0.5) node[below]{$1-\eta_1$} ;
  \draw [<->, dotted] (2,0.3) -- (3,0.3);
  \node at (2.5,0.5) {$\eta_2(1-\eta_1)$};
  \draw [<->, dotted] (3,0.3) -- (5,0.3);
  \node at (4,0.5) {$(1-\eta_2)(1-\eta_1)$};
\end{tikzpicture}
\caption{The stick-breaking process}
\label{fig:stick-breaking}
\end{figure}

Alternative definitions of DPs are also useful both for their understanding and simulation. In particular it can be proved that a \textit{Dirichlet Process} $G \sim \mathrm{DP}(\alpha, H)$, can also be defined via the \textit{stick-breaking} constructive approach \cite{sethuraman1994}. The idea is to build a discrete distribution by assigning probabilities $\pi_k$ to samples $\theta_k$ drawn independently from $H$. As the probabilities $\pi_k$ must sum to 1, a unit-length stick is divided as displayed on Figure \ref{fig:stick-breaking}. The stick is first broken into two parts, of lengths $\eta_1$ and $1-\eta_1$. Then the second portion, of length $1-\eta_1$, is broken again into two parts in proportions $\eta_2$ and $1-\eta_2$. The three resulting portions are now of lengths $\eta_1$, $\eta_2(1-\eta_1)$ and $(1-\eta_2)(1-\eta_1)$. The process of breaking the stick into smaller parts continues indefinitely.

 The weights $\pi_k$ are defined as $\pi_1 = \eta_1$, $\pi_2 = \eta_2 (1-\eta_1)$, $\pi_3 = \eta_3 (1-\eta_2)(1-\eta_1)$, and in general $\pi_k = \eta_k \prod_{l=1:k-1}(1-\eta_l)$. The proportions $\eta_k$ are independent and $\eta_k \sim {\text{\bf Beta}}(1,\alpha)$, where ${\text{\bf Beta}}(a,b)$ is the beta distribution with parameters $a$ and $b$ and probability density function $x^{a-1}(1-x)^{b-1}\un_{[0,1]}(x)$. 
 
 The distribution with weights ${\bm \pi}=[\pi_1,\pi_2,\ldots]$ is called a Griffiths-Engen-McCloskey distribution, denoted by ${\bm\pi}\sim{\text{\bf GEM}}(\alpha)$. Clearly, $\sum_{k=1:\infty}\pi_k=1$. We then get the stick-breaking representation of the Dirichlet Process $G$:
\begin{equation}
    G = \sum_{k=1:\infty}\pi_k\delta_{\theta_k},\quad{\text{with}}\quad{\bm\pi}\sim{\text{\bf GEM}}(\alpha)\quad {\text{and}}\quad\theta_k \sim H.
\end{equation}
Note that the $\pi_k$s tend to decay to zero at geometric rate. Indeed it can easily be proven that:
\begin{equation}
    \mathbb{E}[\pi_k]=\mathbb{E}[\eta_k] \prod_{l=1:k-1}(1-\mathbb{E}[\eta_l])=\frac{1}{\alpha+1}\left(\frac{\alpha}{\alpha+1}\right)^{k-1}.
\end{equation}

Now, suppose we want to fit a mixture model to some observations $y_{1:T} = (y_1, y_2, \ldots, y_T)$. Assume that the mixing distributions are in the form $p_{\theta}(y)$, where $\theta$ is a vector of parameters and that the prior distribution over the vector of parameters is $\theta\sim H$. We can build a nonparametric Bayesian generative model of observations in the form of a Dirichlet Process Mixture model (DPMM). In this model the distribution of observations is a mixture:
\begin{equation}
p(y)=\sum_{k=1:\infty}\pi_k\ p_{\theta_k}(y)
\end{equation}
and the weights $\pi_k$ and parameters $\theta_k$ of the different components of the mixture are defined as a Dirichlet Process: 
\begin{equation}
G = \sum_{k=1:\infty}\pi_k\delta_{\theta_k} \sim DP(\alpha,H)
\end{equation}

\subsection{Hierarchical Dirichlet Process HMM}

The idea of using a DP as a prior in mixture models has been extended to the case of Hidden Markov Models (HMMs). In fact, for some technical reasons that we will explain, the extension of this approach to HMM modelling involves a hierarchy of DPs. 

In the Hierarchical Dirichlet Process HMM (HDP-HMM), DPs are used as priors on the rows ${\bm \pi}_i=(\pi_{i1}, \pi_{i2}, \ldots, \pi_{ik}, \ldots)$ of the transition matrix $\Pi$ of the hidden Markov chain $(z_t)_t$. This makes it possible to specify that the number of states of the Markov chain is unknown.

But it is also necessary to ensure that the transition probabilities $\pi_{ik}$, for all row $i$, weight the same emission distribution $p_{\theta_k}$. This is made possible by parameterizing the DPs  $G_i$ ($i=1,2,\ldots$) by the same \textit{discrete valued} base distribution $G_0$
\begin{equation}
G_i = \sum_k \pi_{ik} \delta_{\theta_k} \sim DP(\alpha, G_0)
\end{equation}
where $G_0$ is modeled by a DP prior with base distribution $H_{\lambda}$: \begin{equation} G_0 = \sum_k \beta_k \delta_{\theta_k} \sim DP(\gamma, H_{\lambda}) \end{equation}

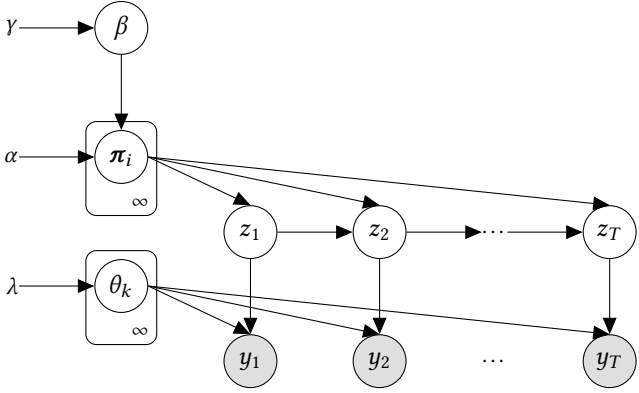
\begin{figure}
\begin{center}
\begin{tikzpicture}
    \node[latent]                (beta)    {$\beta$};
    \node[latent, below=of beta] (pi_k)    {$\bm \pi_i$};
    \node[latent, below=of pi_k] (theta_k) {$\theta_k$};
    
    \node[const, left=of beta]    (gamma)  {$\gamma$};
    \node[const, left=of pi_k]    (alpha)  {$\alpha$};
    \node[const, left=of theta_k] (lambda) {$\lambda$};
    
    \edge {gamma}  {beta};
    \edge {beta}   {pi_k};
    \edge {alpha}  {pi_k};
    \edge {lambda} {theta_k};
    
    \plate {} {(pi_k)}    {$\infty$};
    \plate {} {(theta_k)} {$\infty$};
    
    \node[latent, right=of pi_k, yshift=-1cm] (z1) {$z_1$};
    \node[latent, right=of z1] (z2) {$z_2$};
    \node[const,  right=of z2] (zd) {\ldots};
    \node[latent, right=of zd] (zT) {$z_T$};
    
    \node[obs,   below=of z1, right=of theta_k, yshift=-1cm] (y1) {$y_1$};
    \node[obs,   below=of z2, right=of y1] (y2) {$y_2$};
    \node[const, below=of zd, right=of y2] (yd) {\ldots};
    \node[obs,   below=of zT, right=of yd] (yT) {$y_T$};

    \edge {z1} {z2};
    \edge {z2} {zd};
    \edge {zd} {zT};
    
    \edge {z1} {y1};
    \edge {z2} {y2};
    \edge {zT} {yT};
    
    \edge {pi_k.east}    {z1.north,z2.north,zT.north};
    \edge {theta_k.east} {y1.north,y2.north,yT.north};
\end{tikzpicture}
\end{center}
\caption{The Hierarchical Dirichlet Process - Hidden Markov Model (HDP-HMM).}
\label{fig:pgm-hdp-hmm}
\end{figure}

This hierarchy of DPs yields the HDP-HMM process \cite{teh_hierarchical_2006}. A graphical representation of the HDP-HMM is given in Figure \ref{fig:pgm-hdp-hmm}, where the arrows represent the dependencies. The HMM  itself is represented by states $z_t$ and observations $y_t$. Its parameters are $(\theta_k)_{k\geq 1}$ and $({\bm \pi}_i)_{i\geq 1}$, where $p_{\theta_k}(y_t)=p(y_t\vert z_t=k)$ and ${\bm \pi}_i$ denotes the $i^{\text th}$ row of the transition matrix $\Pi$ of the HDP-HMM, so $\pi_{ij} = \Pprob(z_{t+1}=j \mid z_t=i)$. 

$\alpha,\gamma$ and $\lambda$ are hyper-parameters. $\gamma$ and $\lambda$ are the parameters of a Dirichlet process $G_0\sim DP(\gamma,H_{\lambda})$ that lies at the top of the HDP hierarchy. These random dependencies and vague priors introduce enough flexibility in the model to let it adapt to many different time series.

\color{black}

\subsection{Inference in DP mixtures}
\label{sec:inference-dpm}

Inference in DPMMs is better addressed via the so called Polya urn representation of DPs than through stick breaking. Imagine an urn that contains black and colored balls. The "values" of balls are their colors. At initialization the urn only contains $\alpha$ black balls. When drawing a ball from the urn, if the ball drawn is black then a new colored ball is drawn from a base distribution $H$ and the black and colored balls are put back into the urn. If it is not black, the ball is put back into the urn together with a new one of the same color. 
The labels of the infinite sequence of draws follow a DP. 

We are going to use this formalism in a DPMM, where $z_t$ denotes the hidden state and $y_t$ is the observation. The Polya urn model can be described as follows. Let us introduce $\theta'_t = \theta_{z_t}$ the value of $\theta$ associated to $z_t$. If $z_t = k$ then $\theta'_t = \theta_k$ and $y_t$ is distributed according to $p_{\theta_k}(\bullet)$. Given a sequence of random variables $(\theta'_t)_{t>0}$ with 
\begin{equation}
\begin{array}{ll}
                  P(\theta'_1\in B)&=H(B),\quad\text{ and } \\
P(\theta'_{t+1}\in B\vert \theta'_{1:t})&=\dfrac{1}{\alpha+t}\left( \sum_{\tau=1:t}\delta_{\theta'_{\tau}}(B)+\alpha H(B)\right).   
\end{array}
\end{equation} 
it has been shown in \cite{blackwell1973} that the distribution of $\theta'_t$ converges almost surely to a DP($\alpha$,$H$) (when $t \rightarrow \infty$). 

The estimation of the parameters and states of a nonparametric mixture model from the posterior distribution $p(z_{1:T},\theta_{1:K_T}\vert y_{1:T})$, where $y_{1:T}$ represent the data, can be addressed via Gibbs sampling \cite{Neal00markovchain}. The principle of Gibbs sampling \cite{Robert98montecarlo} is to sequentially update, in turn, the values of $z_t, t=1,\ldots$ and $\theta_k, k=1, \ldots$ conditionally to $y_{1:T}$ and to the current value of the other parameters. It requires knowing the distribution of each latent variable conditionally to the observations $y_{1:T}$ and the other latent variables.

Going back to the Polya urn's model, let us index by $1,\ldots,K_t$ the distinct colors of the balls present in the urn at time $t$ and let $z_t$ denote the color index of the new ball. As the role of the balls can be exchanged, letting $z_{-t}=\{z_{1:t-1},z_{t+1:T}\}$ and $n_{-t,k}=\# \{z_{\tau}\in z_{-t};\;z_{\tau}=k\}$ be the number of occurrences of the value $k$ among $z_{-t}$, it can be shown  \cite{Neal00markovchain} that:
\begin{equation}
\label{Eq:Polya}
P(dz_{t}\vert z_{-t})  =\dfrac{1}{\alpha+T-1}(\sum_{k=1}^{K_{-t}}n_{-t,k}\delta_k(dz_{t})+\alpha\delta_{{\small K_{-t}+1}}(dz_t))
\end{equation}
where $K_{-t}$ is the number of distinct elements in $z_{-t}$ with indexing set from $1$ to $K_{-t}$. Equation \ref{Eq:Polya} can be interpreted as follows: knowing the values of $z_{1:t-1}$ and $z_{t+1:T}$, the distribution of $z_t$ is a mixture of the values $k \in z_{-t}$ and of a new index value $(K_{-t}+1)$. The respective weights of this mixture are $\frac{n_{-t,k}}{\alpha+T-1}$ for any $k \in z_{-t}$ and $\frac{\alpha}{\alpha+T-1}$ for the value $K_{-t}+1$.

It can be proven \cite{Neal00markovchain} that, if observations $y_{1:T}$ and parameters $\theta'_{-t}$ are moreover taken into account, it comes that:
\begin{equation}
\label{Eq:Neal}
\begin{array}{l}
P(dz_{t}\vert z_{-t},y_{1:T},\theta'_{-t})\\ 
\hspace{.3cm}\propto \sum_{k=1}^{K_{-t}}n_{-t,k}p_{\theta_k}(y_t)\delta_k(dz_{t}) 
+\alpha\mathcal{I}(y_t)\delta_{{\small K_{-t}+1}}(dz_t).
\end{array}
\end{equation}
where $\mathcal{I}(y_t) = p(y_t \mid z_t = k, \theta'_{-t}) = \int p_{\theta}(y_t) H_{\lambda}(d\theta)$. 

Note that, provided $\mathcal{I}(y_t)$ is known,  the proportionality factor in Eq. (\ref{Eq:Neal}) can be obtained from the normalization condition $\sum_k P(z_t = k \mid z_{-t}, y_{1:T}, \theta'_{-t}) = 1$. If $p_{\theta}$ and $H_{\lambda}$ are conjugate distributions, $\mathcal{I}(y_t)$ can easily be calculated in closed form. In other cases one can resort to Metropolis-Hastings simulation using the prior distribution of $z_t$ in (\ref{Eq:Polya}) as an auxiliary distribution \cite{Neal00markovchain} to calculate $\mathcal{I}(y_t)$.

After sampling $z_t, t=1:T$ from Eq. (\ref{Eq:Neal}), $\theta_k, k=1,\ldots$ can be sampled from the following distribution \cite{Neal00markovchain}:
\begin{equation}
P(d\theta_k\vert z_{1:T},y_{1:T},\theta_{-k}) \propto  H_{\lambda}(d\theta_k) \prod_{\{t;z_t=k\}}p_{\theta_k}(y_t).
\end{equation}
Here again simulation can be performed directly or via Metropolis-Hastings simulation depending whether $p_{\theta}$ and $H_{\lambda}$ have conjugate distributions. 

\subsection{Inference in HDP-HMMs}
\label{sec:inference-hdphmm}

Inference in HDP-HMM is technically more involved than for mixture models. We briefly summarize it here. Interested readers can find additional information in appendices of \cite{fox2011sticky}. 

Letting $K$ denote the current number of states, the Gibbs sampler should sample $z_{1:T}$. Note that $\theta_{1:K}$ can be marginalized out and does not need to be sampled in Gibbs. To make it possible, we will also have to sample the $\pi_j$, which in turn requires sampling the weights of the base distribution $G_0=\sum_{k=1:\infty}\beta_k\delta_{\theta_k}$. As only $(\beta_k)_{k=1:K}$ is concerned for describing the weights of the states of the finite size data set at hand, letting $\beta_{-K} = \sum_{k=K+1:\infty}\beta_k = 1-\sum_{k=1:K}\beta_k$, we simply sample $(\beta_{1:K},\beta_{-K})$ that follows a Dirichlet distribution of order $K+1$. The sampling of $(\beta_{1:K},\beta_{-K})$ is described in \cite{fox2011sticky}. 

Note also that we want to implement inference for a sticky HDP-HMM, that is, a modified version of the HDP-HMM that models persistency of the states by biasing the model towards self transitions $(z_{t-1}=j,z_t=j)$. This is ensured by introducing an additional parameter $\kappa$ and changing the prior upon $\pi_j$:
\begin{equation}
\pi_j\vert \alpha,\beta,\kappa\sim {\text{\bf DP}}
\left(\alpha+\kappa,\dfrac{\alpha(\sum_k\beta_k\delta_{k})+\kappa\delta_j}{\alpha+\kappa}\right).
\end{equation}
When $\kappa=0$ we get the standard HDP-HMM, while when $\kappa\rightarrow\infty$, $\pi_j$ tends to only weight state $j$. 

To implement the Gibbs sampler for the states $z_{1:T}$ let $\psi=(\alpha,\beta,\kappa,\lambda)$, and $\bm{\pi} = (\pi_j)_j$. Then $P(z_t\vert y_{1:T},z_{-t},\psi)$ can be expressed by marginalizing against the ${\bm \pi_j}$s and $\theta_k$s:

\begin{equation}
\label{eq:hdp-hmm_Gibbs_z_t}
P(z_t \,|\, y_{1:T},z_{-t},\psi)  \propto P(z_t\,|\,z_{-t},\psi) p(y_t\,|\,y_{-t},z_{1:T},\psi)
\end{equation}

Let us introduce the following notations: $x_{i\bullet}=\sum_j x_{ij}$ and $n_{jk}^{-t}$ denotes the number of transitions from state j to state $k$, not counting the transitions $z_{t-1} \rightarrow z_t$ or $z_t \rightarrow z_{t+1}$. Then, the first factor in (\ref{eq:hdp-hmm_Gibbs_z_t}) writes
\begin{equation}
\begin{array}{l}
    P(dz_t\mid z_{-t},\psi)\\
    \hspace{.3cm}\propto \sum_{k=1}^{K_{-t}}\dfrac{\alpha\beta_k+n_{z_{t-1},k}^{-t}+\kappa\delta_{z_{t-1},k}}
    {\alpha+\kappa+n_{z_{t-1},\bullet}^{-t}}\\
    \hspace{.9cm} \times 
     \dfrac{\alpha\beta_{z_{t+1}}+n_{k,z_{t+1}}^{-t}+\kappa\delta_{z_{t-1},k}\delta_{k,z_{t+1}}}{\alpha+\kappa+n_{k,\bullet}^{-t}+\delta_{z_{t-1},k}} \delta_k(dz_t)\\
    \hspace{.6cm}+ \dfrac{\alpha^2\beta_{-K}\beta_{z_{t+1}}}{(\alpha+\kappa)^2}\delta_{K_{-t}+1}(dz_t)
\end{array}
\end{equation}
The second factor in (\ref{eq:hdp-hmm_Gibbs_z_t}) writes
\begin{equation}
\begin{array}{ll}
    p(y_t\mid y_{-t},z_{1:T},\psi)\\
    \hspace{.3cm}\propto\int_{\theta_{z_{t}}}
    p(y_t\mid \theta_{z_{t}})H_{\lambda}(d\theta_{z_{t}}\mid\{ y_{\tau};z_{\tau}=z_t,\tau\neq t\})
    \end{array}
\end{equation}
As far as discussed earlier, if the $\theta_k$s have conjugate prior distributions, $p(y_t\mid y_{-t},z_{1:T},\lambda)$ can be calculated in closed form.
Note in addition that to avoid a particular choice of hyperparameters $(\alpha,\gamma,\lambda)$ biasing the solution, they can also be given some prior distribution.

At the end of the process, after the $z_{1:T}$ have been estimated, the $\theta_k$s can be estimated easily, e.g. by maximizing the likelihood $p(\{y_t; z_t = k\}\,|\,\theta_k)$.

\subsection{Section summary}
\label{sec:npbayesian-summary}

In this section we have introduced non-parametric Bayesian approaches. In Bayesian statistics some of the parameters on which the data depend are considered random. The term "non-parametric" means that there is a large number of parameters that are estimated from the data. 

When the number of states of a mixture or a HMM is not known in advance, it is possible to use a non-parametric Bayesian approach using Dirichlet processes (DP) as priors. This is called the Dirichlet Process Mixture Model (DP-MM) or the Hierarchical Dirichlet Process Hidden Markov Model (HDP-HMM). Equivalently, the name infinite (or non-parametric) mixture or HMM can also be used. 

Missing data, that is, states and parameters, can be estimated from observations using a Gibbs sampling algorithm which comes up to randomly simulating, in turn, the different components of the model which are not measured directly. These components are simulated according to some conditional distributions which have been specified in this section. 

\section{A first look at RTTs through the HDP-HMM}
\label{sec:first-look}

As stated previously, HDP-HMM is a flexible method for inferring HMM parameters and segmenting data when the number of latent states is unknown. This fits the problem of segmenting RTT time series (remember that of Figure \ref{fig:sample-raw-rtt}), where the number of different states is not a priori known.
Furthermore, it is not mandatory to make an assumption on the type of RTT distribution in each state (Gaussian, exponential ...). This distribution can be assumed nonparametric, which introduces even more flexibility and allows a very generic model that adapts to a very large number of traces.

In this section we show that the model produces realistic segmentations from a human point of view, and that the inferred parameters are easily interpretable with respect to the application domain. In addition, we provide two validations for the model. We show on a labeled change point dataset that the model performs at least as well as ad-hoc change point detection methods. And we also show that the states inferred from the RTT time series match well with the AS and IP paths seen in RIPE Atlas traceroutes.

\subsection{A nonparametric observation model}

Many parametric models have been proposed in the literature to explain the distribution of the delay in computer networks and on the Internet. For example, in \cite{Sato2005} a Gaussian mixture model is proposed, in \cite{Hernandez2006} a Weibull mixture model, and in \cite{Zhang2007} a Pareto distribution. In practice, however, it seems that the distribution can be very different depending on the network state. For example, in some states the delay can be relatively stable with occasional spikes above a baseline, in which case it might be modeled by an exponential distribution, while in other states the delay can experience large variations caused by a high traffic level, and might be better modeled by a normal distribution.

In this work we choose instead to use a nonparametric Dirichlet Process Mixture Models (with a Gaussian as "base" distribution) as emission distributions of the HDP-HMM. As such, the delay in each state is modeled by a varying number of Gaussian components. This allows us to model a wide range of distributions, and we avoid choosing a particular parametric emission distribution for each state of the HDP-HMM. For each Gaussian component, we use a Normal-Inverse-$\chi^2$ prior, which is the conjugate prior to the normal distribution with unknown mean and variance.
The use of a nonparametric observation model reinforces the need for Bayesian inference methods, since a more traditional MLE approach would require several layers of penalization. 

\begin{figure}[h]
    \centering
    \includegraphics[width=\linewidth]{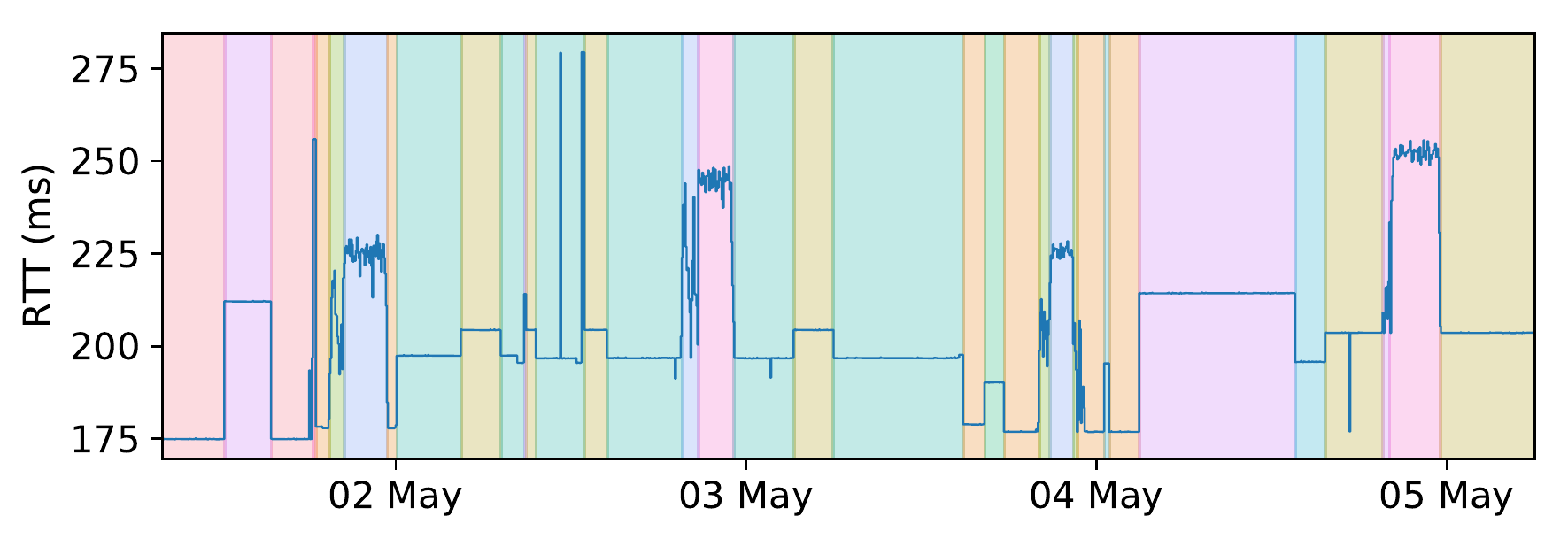}
    \caption{Segmentation of RTT observations between \texttt{at-vie-as1120} and \texttt{sg-sin-as59253}.}
    \label{fig:sample-segmented-rtt}
\end{figure}

The segmentation of the series from Figure \ref{fig:sample-raw-rtt} using an HDP-HMM with DP-GMM emissions is shown in Figure \ref{fig:sample-segmented-rtt}. A same state corresponds to a same color.

\begin{figure*}[h]
    \centering
    \includegraphics[width=\textwidth]{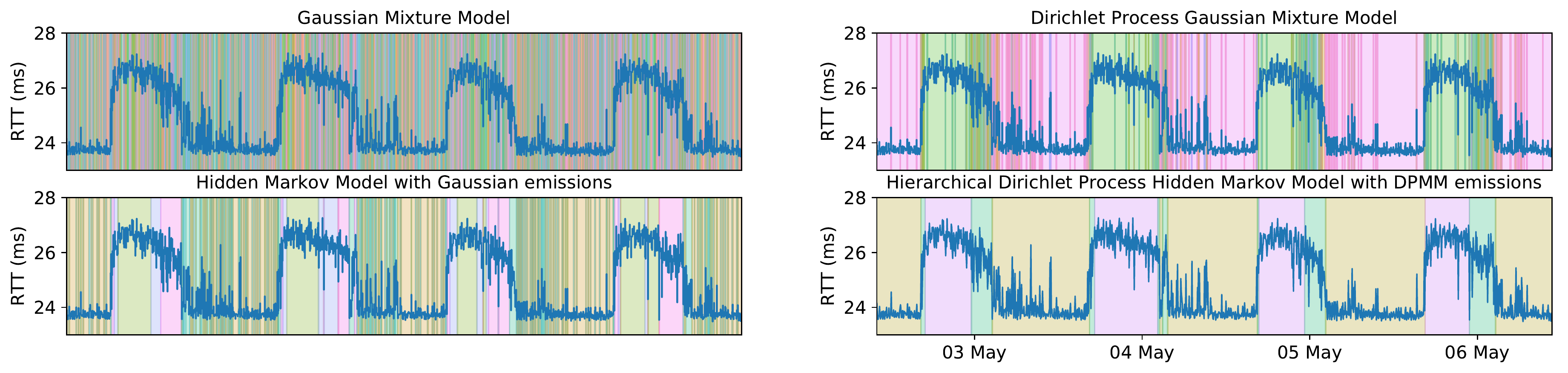}
    \caption{Segmentation of a RTT time series with parametric and nonparametric mixture models and HMMs.}
    \label{fig:models-comparison}
\end{figure*}

As a matter of comparison, we provide in Fig. \ref{fig:models-comparison} the segmentation obtained with a HDP-HMM with DPMM emission distributions, with that resulting from parametric and nonparametric MMs and HMMs with a Gaussian observations model. In the case of the gaussian MM and of the HMM, the number of latent states has been chosen by estimating the model for a varying number of components and choosing the number that minimizes the penalized log-likelihood using the BIC criterion. As we can see the HDP-HMM produces a segmentation close to what a human would do, contrary to other models which generate far too many state changes.

\subsection{Change point detection}
\label{sec:cpt}

Quantifying the performance of the HDP-HMM on real RTT time series is not easy since there is no ground truth. The "network state" is not known or vaguely defined. But it was possible to compare the performance of the model on a change point detection task where the goal is to detect significant changes in the delay. While not the primary purpose of the HDP-HMM, detecting change points is simply a matter of segmenting the data and finding changes in the inferred state sequence and this this allows to partially validate the quality of the segmentation obtained.

We have benchmarked the HDP-HMM against different change point detection methods on a labeled dataset introduced by \cite{Shao2017}. This dataset is particularly interesting because change points in RTT timeseries have been manually labeled by human experts. To our knowledge, there are no other RTT time series datasets that are both realistic and labeled. 

The dataset consists of 50 RTT series of varying length for a total of 34,008 hours of observations. In \cite{Shao2017} change point detection is performed by minimizing $\sum_{i=1}^{m+1} C(y_{\tau_{i-1}+1:\tau_i}) + \beta f(m)$. $m$ is the number of changes, $C$ is a cost function that measures the stability of the delay over a range of successive values, and $f(m)$ is a penalty that prevents overfitting. Different cost functions and penalties are considered.

We have compared the performance of the segmentation obtained by HDP-HMM with the best performing changepoint detection methods of \cite{Shao2017}. In our approach a HDP-HMM model is learnt on each timeseries, the most likely hidden state sequence is computed, and changepoints are simply defined as changes in the hidden states sequence.

\begin{figure*}
    \centering
    \includegraphics[width=\textwidth]{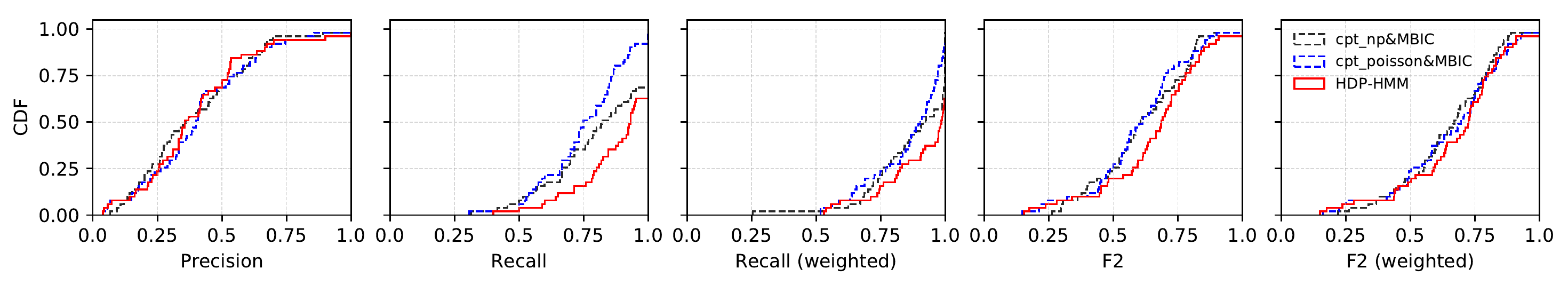}
    \caption{Benchmark of the HDP-HMM against classical change point detection methods on a human-labelled change point dataset \cite{Shao2017}. The weighted recall gives more importance to large delay changes.}
    \label{fig:benchmark-cpt}
\end{figure*}

We show on Fig. \ref{fig:benchmark-cpt} that the HDP-HMM performs similarly to the best performing change point detection methods of \cite{Shao2017} in terms of precision ($\frac{\text{\# True Positive}}{\text{\# True Positive}+\text{\# False Positive}}$), while performing better in terms of recall ($\frac{\text{\# True Positive}}{\text{\# True Positive}+\text{\# False Negative}})$. This means that our model is more sensitive to small changes in the delay without generating unnecessary false alarms.

\subsection{RIPE Atlas measurements}

In addition to detecting significant changes in the delay, the HDP-HMM also provides a notion of hidden states. In this section we validate the quality of this clustering visually and by studying the correlation with AS and IP paths revealed by traceroutes.

\subsubsection{Dataset}

RIPE Atlas offers two types of measurement sources: probes and anchors. Probes are deployed in heterogeneous environments while anchors are restricted to high-availability environments such as data centers, universities, and IXPs (Internet eXchange Points). Anchors tend to be located in well-connected autonomous systems and measurements between anchors represent more stable paths than what may be observed from probes located at the edges of the Internet. On the other hand, anchors are more powerful and perform the so-called \emph{anchoring mesh measurements}, where various measurements are performed regularly between each pair of anchors. This allows us to collect traceroute results both on the forward and on the reverse path.

Our dataset consists of one week of IPv4 RTT measurements between all Atlas anchors and the \texttt{at-vie-as1120} anchor\footnote{RTT measurements results are available at \url{https://atlas.ripe.net/measurements/1437285}. We considered the period between the 2nd and the 9th of May 2018.}. Delay is measured every four minutes using three ICMP (Internet Control Message Protocol) pings towards the target anchor. We kept the minimum value of the delay for each time step. Considering the subset of anchors that were online over the time period, we collected 301 series of 2520 data points. We also collected the associated traceroute measurements, both on the forward path, and on the reverse path. Traceroutes are performed every fifteen minutes using three ICMP probe packets for each hop.

\begin{figure*}
    \centering
    \includegraphics[width=\textwidth]{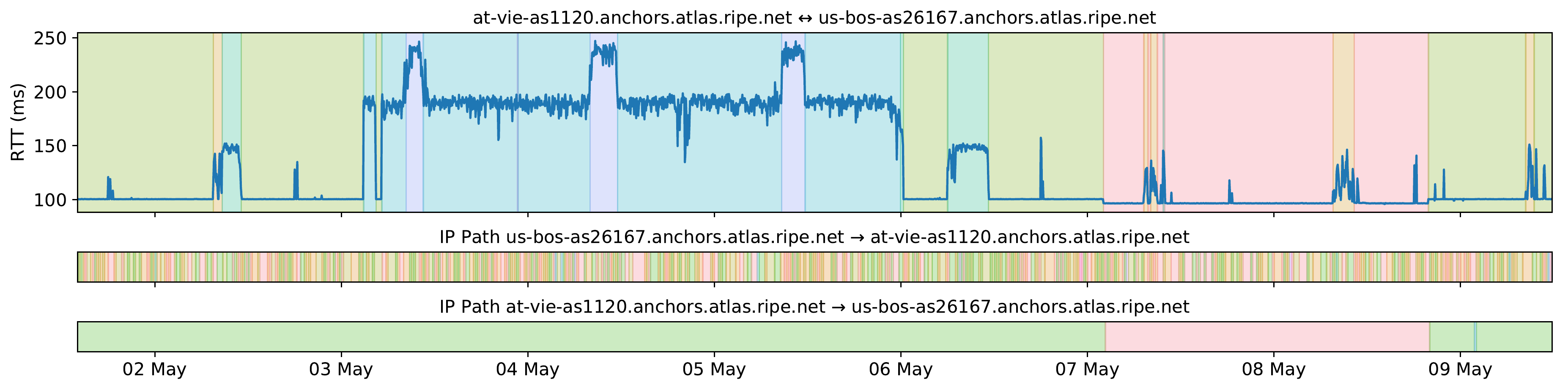}
    \caption{Segmentation of RTT observations between \texttt{at-vie-as1120} and \texttt{us-bos-as26167} using an HDP-HMM with DP-GMM emissions. Each color identifies a state or an IP path observed in the traceroute.}
    \label{fig:example-path}
\end{figure*}

\subsubsection{Inference}

We have segmented each series using our Julia implementation of the Gibbs sampler. It takes less than 5 seconds on a single thread of a 2.80GHz Intel Core i7-7600U CPU to process a 2520 point time series (1 week of an Atlas RTT measurement) with 500 iterations of the sampler. The task is highly parallelizable as each time series can be processed independently. Using 4 threads, 300 one-week long time series can be processed in 6 minutes.

\begin{figure}[t]
    \centering
    \includegraphics[width=\linewidth]{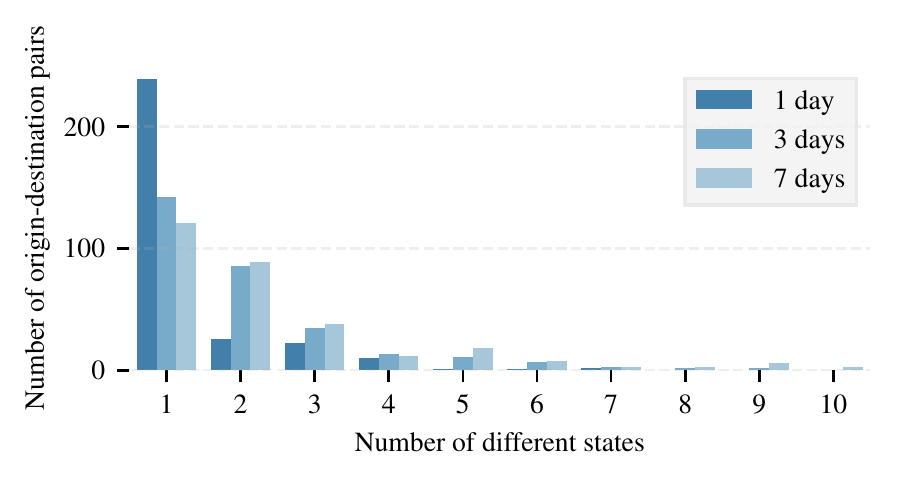}
    \caption{Distribution of the number of states learned for different timescales.}
    \label{fig:hmm-states-dist}
\end{figure}

Figure \ref{fig:hmm-states-dist} shows the distribution of the number of states in the resulting HMMs for different measurement timescales. It is clear that the number of states grows with the length of the series. This is not surprising and visual inspection by a human expert would also probably identify more states in longer timeseries. One, three, and seven days long series have respectively less than 8, 10, and 11 states. This confirms the capability of the HDP-HMM to learn more complex models as the number of RTT observations, and possibly the number of underlying network configurations, grows.

\subsubsection{State durations vs. delay variations}

An advantage of HMMs over other timeseries models (e.g. autoregressive models or neural networks) is that the parameters are more easily interpretable with respect to the application domain. In our case, the state transition matrix ${\bf \Pi}$ gives us information about the frequency of network configuration changes and the relation between them, while the observations distributions give us in particular the mean value of the delay and its variance (of the delay in each configuration).

On most time series we can distinguish two types of states: those where the delay is relatively constant (such as the green one on Fig. \ref{fig:example-path}), and states where the delay is very variable (such as the purple one). This is reflected by the variance of the delay in the state. 
And the average duration of a HMM in a state $i$ is given by $1/(1-\pi_{ii})$ where $\pi_{ii}$ is the probability of self-transition. In the example of Fig. \ref{fig:example-path} the average duration of the purple state is of 45 timesteps (= 3 hours) and of 149.5 timesteps (= 9 hours 58 minutes) for the green state. The standard deviation of the delay in the purple state is of $\sigma = 10.3$ msec while the standard deviation of the green state if of $\sigma = 4.1$ msec. States with a high variance could possibly be explained by intra-domain load-balancing (since Atlas pings flow ID is not constant), congestion, or in-path devices delaying the processing of ICMP packets. However asserting the cause of such variations and studying the possibility of detecting them from delay measurements is to be done in future works.

\begin{figure}[t]
    \centering
    \includegraphics[width=\linewidth]{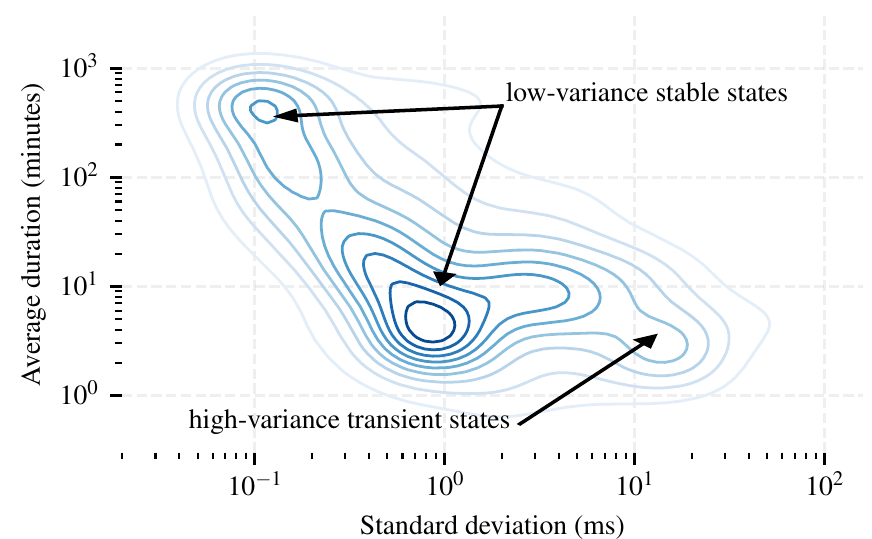}
    \caption{Density estimation of the (standard deviation, average duration) couple. Darker colors indicate a higher density.}
    \label{fig:hmm-std-vs-duration}
\end{figure}

Figure \ref{fig:hmm-std-vs-duration} displays the standard deviation of the RTT against the average duration in a state. In the analyzed dataset the average state duration decreases as the RTT standard deviation increases. This is not surprising as we expect Internet paths to spend more time in stable states.

\subsubsection{Relationship with the AS and IP paths}

\begin{figure}[t]
    \centering
    \includegraphics[width=\linewidth]{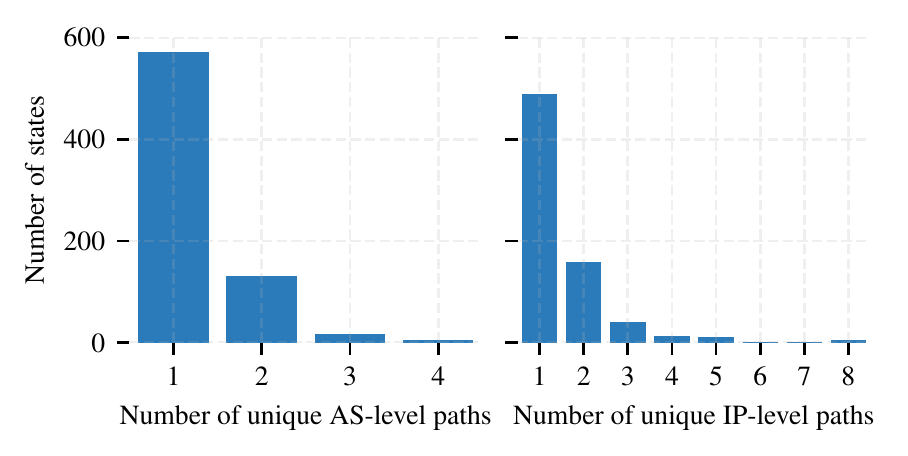}
    \caption{Distribution of the number of states associated with a given number of unique paths.}
    \label{fig:states-paths-counts}
\end{figure}

We hypothesized that the distribution of delay observations is conditioned on the underlying network state, such as the inter and intra-AS routing configuration, as well as the traffic level. As shown in Figure \ref{fig:states-paths-counts}, the majority of the states learned over all the paths in our dataset matches only one AS path and one IP path. For example there are 595 states which always correspond to the same AS path over the 746 states learned. Stated differently only 16\% of the states learned can match two AS paths or more. States associated with more than one AS path can be explained by delay differences too small to be separated into two clusters. 

Conversely, one AS or IP path can be mapped to several states. For example in Figure \ref{fig:example-path} we only observe the AS path \texttt{ASN MARKLEY} $\rightarrow$ \texttt{GTT BACKBONE} $\rightarrow$ \texttt{NTT COMMUNICATIONS} $\rightarrow$ \texttt{ACONET SERVICES} in the traceroutes from \texttt{us-bos-as26167} to \texttt{at-vie-as1120} and \texttt{ACONET SERVICES} $\rightarrow$ \texttt{ACONET} $\rightarrow$ \texttt{NEXTLAYER AS} $\rightarrow$ \texttt{NTT COMMUNICATIONS} $\rightarrow$ \texttt{ASN MARKLEY} in the reverse traceroutes (as resolved using the RIPEstat API). In the forward traceroutes we observe IP path changes every 15 minutes, in the GTT and NTT ASes, probably due to intra-AS load-balancing, while in the reverse traceroutes we only observe two different IP paths in NTT AS that are perfectly correlated to state changes in the model.

\subsection{CAIDA MANIC measurements}

In addition to RIPE Atlas delay measurements, the HDP-HMM fits other kinds of network measurements as well. In this section we show the results obtained on delay measurements from the CAIDA MANIC project \cite{Manic}.
The CAIDA MANIC project uses Time Series Latency Probes (TSLP) to measure inter-domain congestion. Once a peering link between two ASes has been identified, ICMP probes are sent to the near-end (i.e. the last router in the first AS) and the far-end (i.e. the first router in the second AS) of the link. The intuition is that if there is congestion the router queues will fill up, and the delay between the near-end and the far-end will increase. Using the same model as for the RIPE Atlas RTT series, we segment the delay difference time series (far-end - near-end) from publicly available measurements. 

\begin{figure*}
    \centering
    \includegraphics[width=\textwidth]{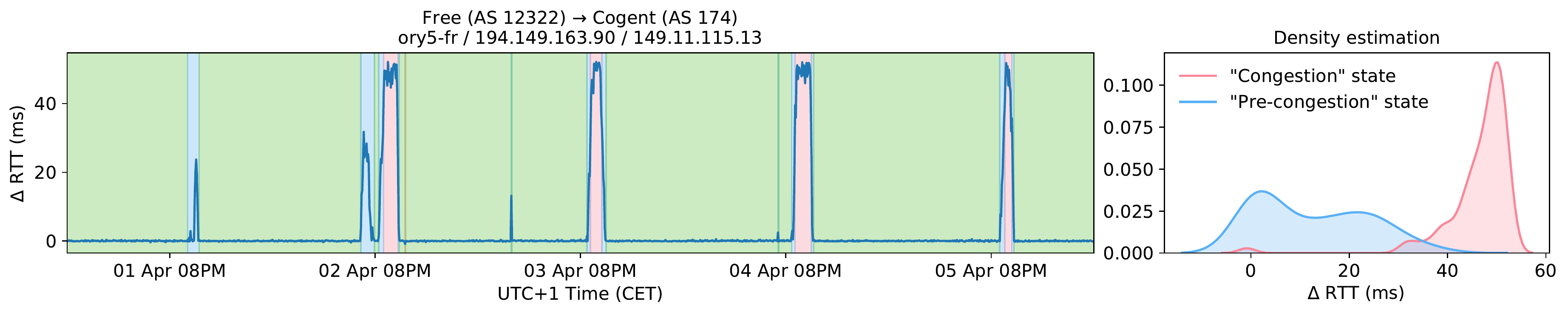}
    \caption{Segmentation of a RTT difference (far - near) time series obtained with TSLP probes from the CAIDA MANIC project. Each color identifies a state.}
    \label{fig:segmentation-manic}
\end{figure*}

In Figure \ref{fig:segmentation-manic} we show the resulting segmentation for a peering link experiencing periodic congestion. Three states are learned. The green state, corresponding to a non-saturated link, has a standard deviation of 0.1 ms, while the standard deviation for the red and blue states are of, respectively, 7 ms and 11 ms. The blue state seems to correspond to a state of increased traffic level, while the red state seems to correspond to a saturated link. Because the model accounts for temporal dependencies, it is able to clearly separate those two states even though their distributions are overlapping.

\section{Large-scale measurement analysis}
\label{sec:large-scale}

Internet monitoring projects such as RIPE Atlas provide a large amount of latency information. Due to its scale RIPE Atlas has a good chance to provide enough information to let detect anomalous latency patterns in important network components, such as IXPs or large transit providers. However, detecting and characterising these anomalies has proven challenging (e.g. the analysis in \cite{Aben2015} took weeks). In this section we will show how aggregating change points learned with the HDP-HMM from a large number of origin-destination pairs is a simple and elegant method to detect and characterise anomalies in key Internet infrastructures.

\subsection{RIPE Atlas Trends API}
\label{sec:Atlas-API}
In order to make our method accessible to many people, we have developed a publicly exposed Web API into RIPE Atlas. Given an origin-destination pair (measurement and probe ID) and a time frame (start and stop time), the \emph{trends} API provides the segmentation of a RIPE Atlas delay measurement. The API offers three endpoints, described in Table \ref{tab:api-endpoints}.

\begin{table}[h]
\caption{Endpoints of the Atlas Trends API.}
\label{tab:api-endpoints}
\begin{tabular}{@{}lll@{}}
\toprule
\textbf{Method} & \textbf{Path}                              & \textbf{Parameters} \\ \midrule
GET             & /ticks/:msm\_id/:prb\_id          & start, stop         \\ \midrule
GET             & /trends/:msm\_id/:prb\_id         & start, stop         \\ \midrule
GET             & /trends/:msm\_id/:prb\_id/summary & start, stop         \\ \bottomrule
\end{tabular}
\end{table}

The \texttt{/ticks} endpoint returns the minimum RTT for a given pair with a constant time interval (duplicated results due to probes connectivity problems are suppressed, and missing results are explicitly inserted). The \texttt{/trends} endpoint returns the minimum RTT and the associated segmentation. For example, the URL \url{https://trends.atlas.ripe.net/api/v1/trends/1437285/6222/?start=2018-05-02&stop=2018-05-10} gives the segmentation of the Figure \ref{fig:example-path} (it should take less than 10 seconds to segment one week of data). A summary of the time series, as shown in Listing \ref{lst:api-output}, can also be requested by appending \texttt{/summary} to the path. Start and stop time are specified as \texttt{YYYY-MM-DDTHH:MM} where \texttt{THH:MM} is optional and defaults to the start of the day.

Additionally to this article we provide interactive notebooks to document and demonstrate the API, and compare various statistical models. Links to interactive Google Colab sessions, as well as the notebooks source and code to facilitate the usage of the API are provided on GitHub \cite{trends_demonstration_2019}.

\begin{listing}
\includegraphics[width=\linewidth]{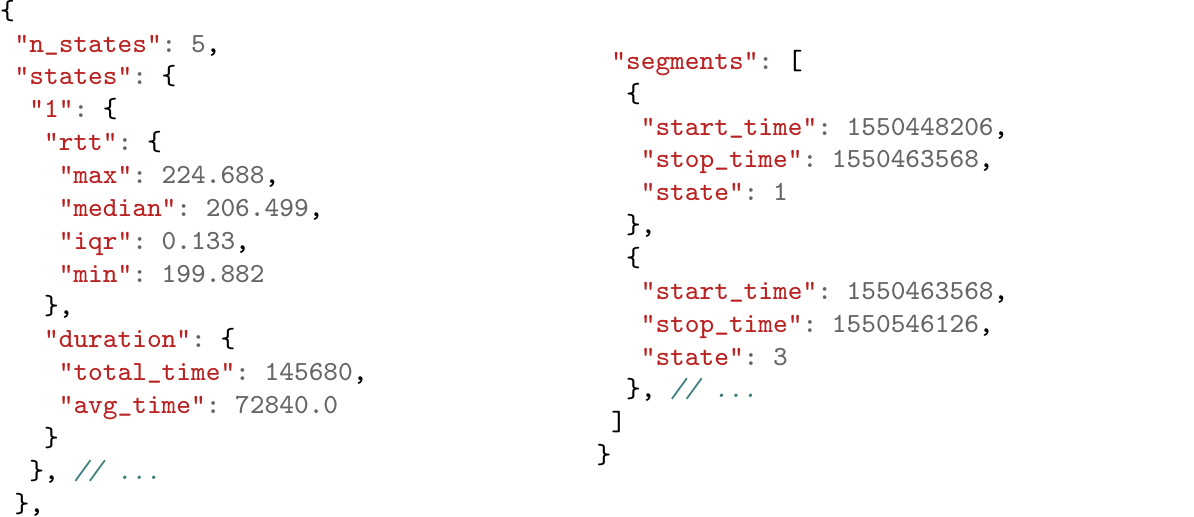}
\caption{RIPE Atlas Trends API sample JSON output.}
\label{lst:api-output}
\end{listing}

\begin{figure*}[htb]
\centering
\begin{minipage}[b]{.48\textwidth}
    \includegraphics[width=\textwidth]{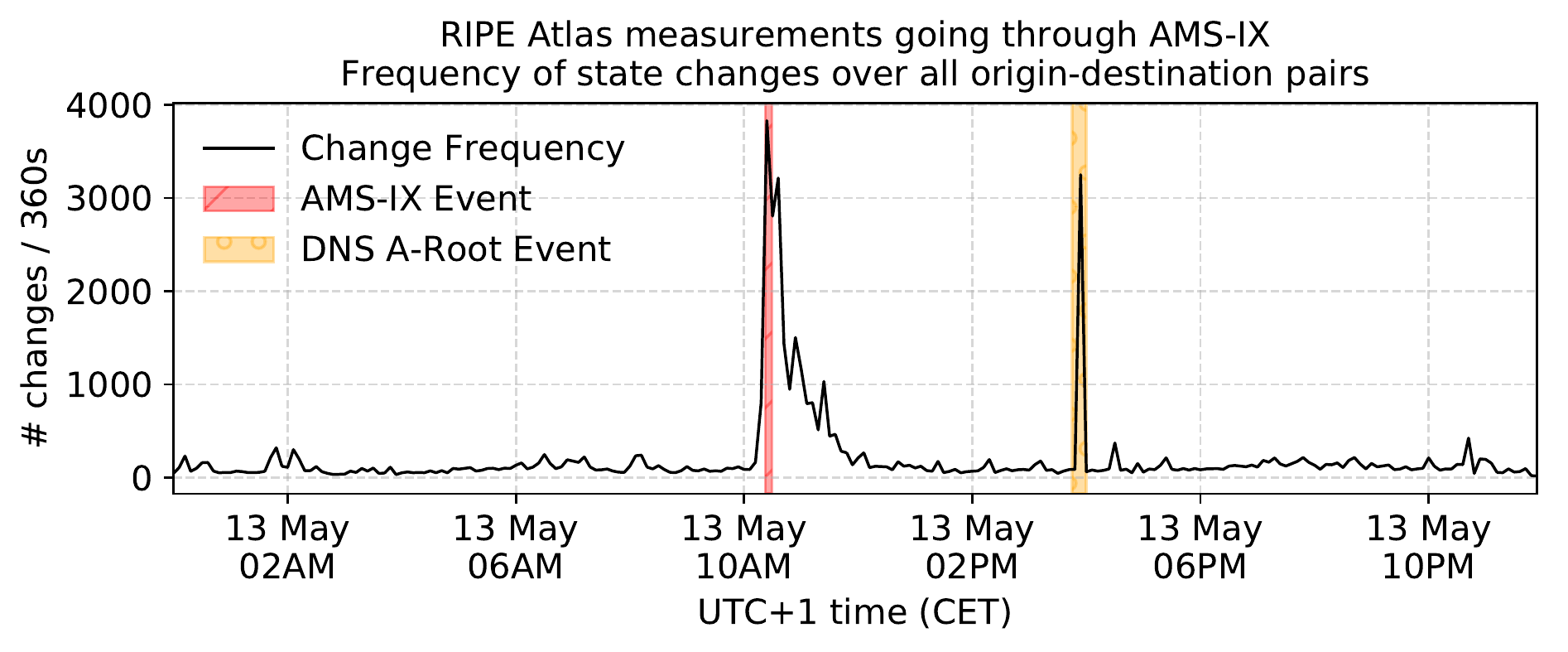}
    \caption{Change frequency between on the 13th of May 2015 for the 20k pairs that saw AMS-IX in their traceroutes the day before.}
    \label{fig:change-freq-ams-ix}
\end{minipage}\qquad
\begin{minipage}[b]{.48\textwidth}
    \includegraphics[width=\textwidth]{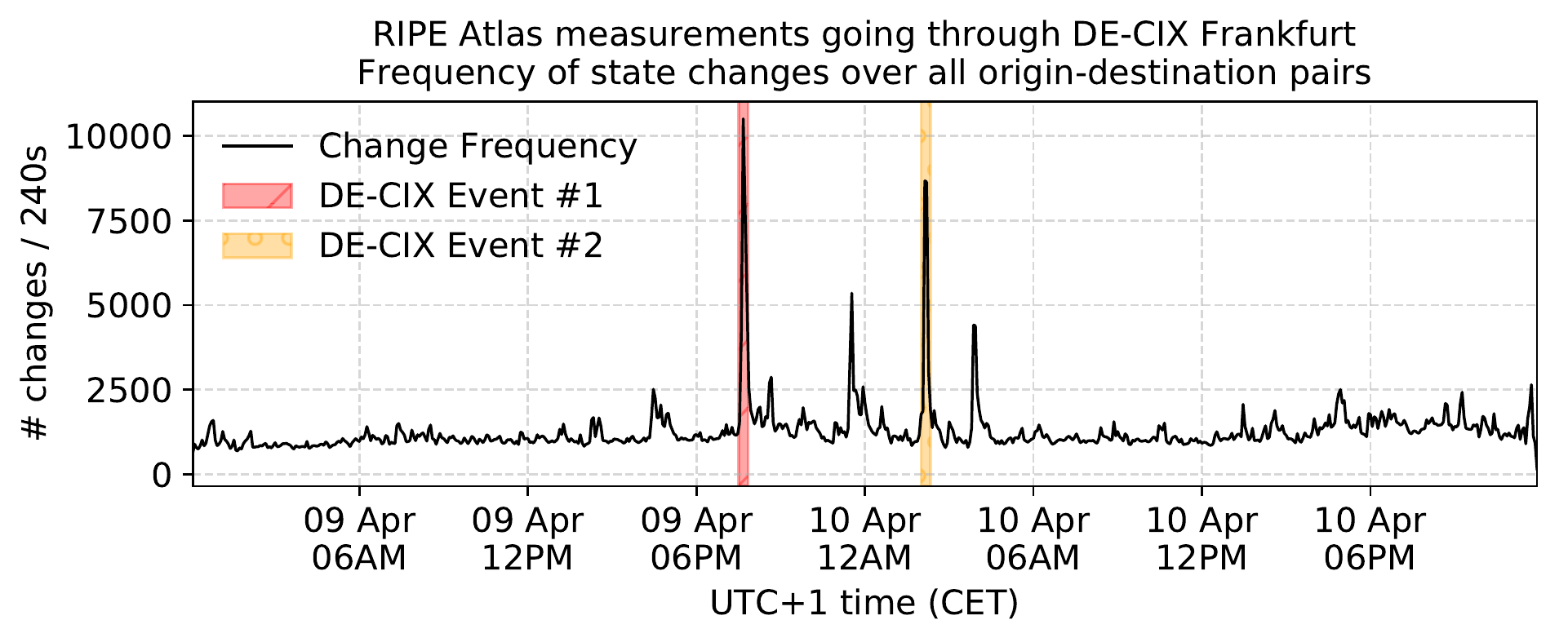}
    \caption{Change frequency between the 9th and the 10th of April 2018 for the 60k pairs that saw DE-CIX Frankfurt in their traceroutes the day before.}
    \label{fig:change-freq-de-cix}
\end{minipage}
\end{figure*}

\subsection{Monitoring large Internet infrastructures}

As shown in \cite{Aben2015}, a significant number of Atlas origin-destination pairs reliably go through large Internet infrastructures, such as IXPs (AMS-IX, DE-CIX, ..) and transit providers (Level 3). By reliably, we mean pairs for which such infrastructures have been seen consistently in the traceroutes over a given time frame. Furthermore, Atlas provides measurements towards the 13 DNS root servers from every probe (more than 10k probes), although such measurements are more difficult to exploit due to the anycast nature of DNS root servers. 

In order to detect anomalous events in those infrastructures, we propose to aggregate the change points learned from each time series individually, to obtain a \emph{state changes frequency} which represents the number of state changes in a given time frame over all the origin-destination pairs considered. One problem is the selection of those origin-destination pairs. One could imagine learning the model for all the origin-destination pairs available in Atlas, or a large subset, such as anchoring mesh measurements (160k origin-destination pairs), and then look for events in the state changes frequency. However, preliminary experimentations show that the obtained signal is too noisy and requires a lot of manual processing to find relevant events. Instead, we propose to monitor each infrastructure individually, by considering only the origin-destination pairs for which the infrastructure has been seen in recent traceroute measurements.

To validate the ability of our method to detect events, we analyze two events which have been discussed in the literature (as this provides some groundtruth against which we can compare our results): AMS-IX outage in May 2015 \cite{Aben2015,Fontugne-IMC2017,giotsas2017}, and DE-CIX Frankfurt outage in April 2018 \cite{Aben2018}.

\subsubsection{AMS-IX May 2015 outage}

According to \cite{Aben2015}, on the 13th of May 2015, AMS-IX experienced a partial outage due to a switch interface generating looped traffic on the peering LAN. The event lasted for seven minutes and two seconds, from 10:22:12 to 10:29:14 (UTC time) before the switch interface was disconnected. This event caused some peers located at AMS-IX to loose their BGP session. In \cite{Aben2015} the event has been studied using traceroutes, by looking at the percentage of paths seeing AMS-IX peering LAN in their traceroute over time. However changes in the IP paths often result in changes in the delay. Using the ping measurements corresponding to the same origin-destination pairs, provided by RIPE NCC, we learned the models and extracted the changepoints.

By default RIPE Atlas ping measurements are performed every 4 minutes, with a jitter of 2 minutes to maximize the temporal coverage over all the probes participating in a measurement. Hence we counted the number of changepoints in buckets of 6 minutes. We show the resulting \emph{state change frequency} on Figure \ref{fig:change-freq-ams-ix}. We highlighted in red the real event duration. The event corresponds to a clearly visible increase in the number of changes. The frequency stays high for a few hours as first of all many peers switch to alternative paths, and then some of them come back to AMS-IX.

We also see a spike between 14:45 and 15:00 (UTC). 
Further investigation has shown that almost all the changepoints that have occured during this period are related to measurements targeting the DNS Root-A server. We have repeated a similar procedure for all the origin-destination pairs in the Atlas built-in measurement to this DNS server and we have seen a similar spike, but all source ASes seem to be affected equally, leading us to believe that the spike was caused by an event close to one of the DNS Root-A instances.

\subsubsection{DE-CIX April 2018 outage}

According to \cite{Aben2018}, between April 9th and April 20th 2018, some networks located at DE-CIX Frankfurt lost their connectivity to the route servers, and as a result rerouted their traffic to other interconnections, or experienced an interruption of traffic. An analysis of the rates of BGP updates received by route collectors located at DE-CIX showed that the rates of updates dropped close to zero between 19:43 and 23:28 on the 9th of April, and between 02:02 and 03:51 on the 10th of April. Applying the same methodology as for the AMS-IX event, we show the state changes frequency for this time frame in Figure \ref{fig:change-freq-de-cix}. The two largest spikes match exactly the two times where the rates of BGP updates dropped to zero. The two smaller spikes match with the two times when the collectors started receiving BGP updates again.

\subsection{Validation of the HDP-HMM model at large scale on RIPE Atlas}

In Section \ref{sec:cpt} we show that the HDP-HMM model is at-least as good as classical change point detection methods on a labelled RTT change points dataset. This however, does not tell us whether the model fits well RTT data from a statistical point of view. In this section, we propose to compare the likelihood of the time series (with respect to their inferred model) with the likelihood of time series simulated according to an HDP-HMM model. 
If the models fit well the data, we can expect that the likelihood of the data with respect to the model should follow the same distribution as the likelihood of synthetic data generated by the model.

To do so, we consider 100k time series of one week duration (2520 data points) from the anchoring mesh measurements.
We learn the model for each time series, and compute their likelihood $p(\bm{y} | \bm{\pi}, \bm{\theta})$ with respect to the model.
In addition, for each HMM with parameters $(\bm{\pi}, \bm{\theta})$ we sample a time series $\bm{y}'$ and compute its likelihood $p(\bm{y}' | \bm{\pi}, \bm{\theta})$.

\begin{figure}[h]
    \centering
    \includegraphics[width=\linewidth]{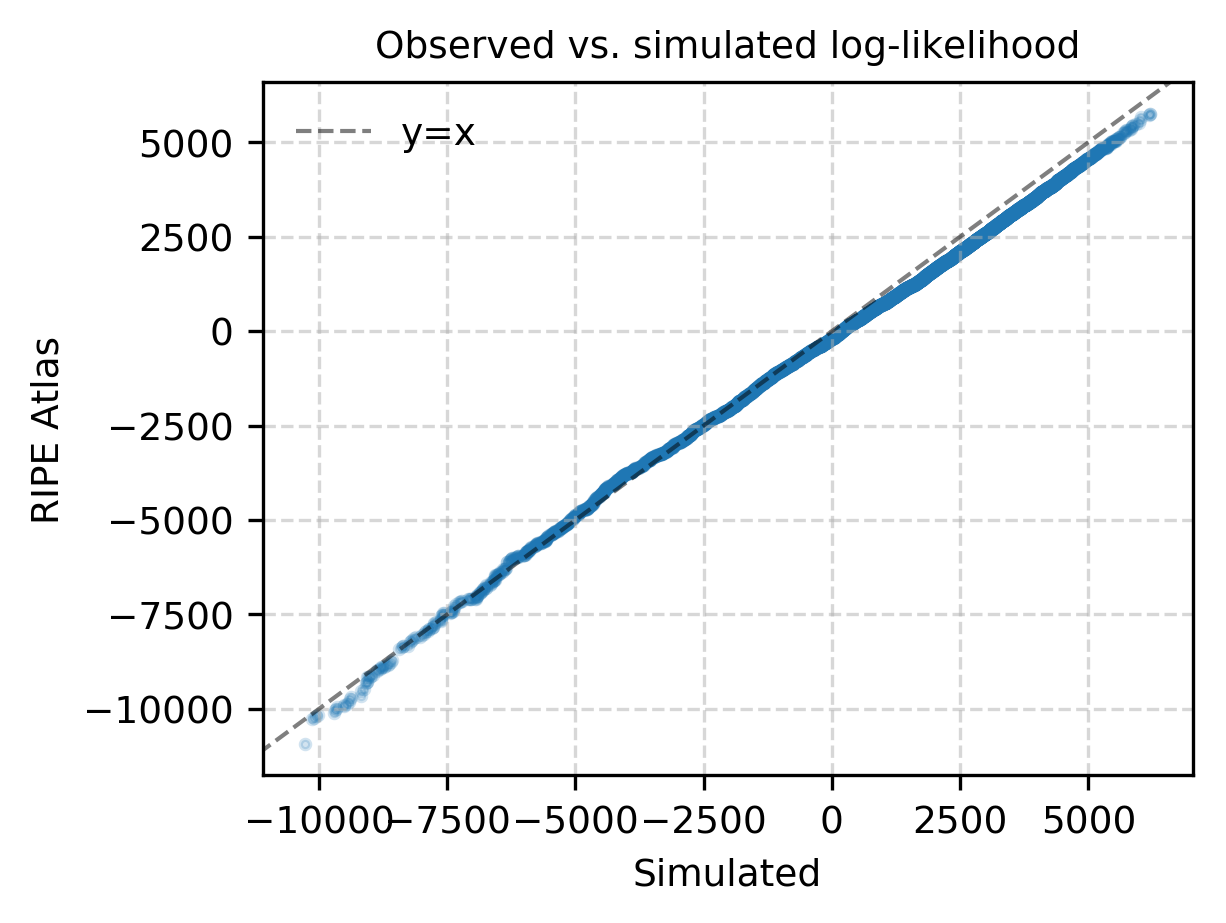}
    \caption{Q-Q plot of observed vs. simulated log-likelihood on 100k time series.}
    \label{fig:log-likelihood-qq}
\end{figure}

\begin{figure}[h]
    \centering
    \includegraphics[width=\linewidth]{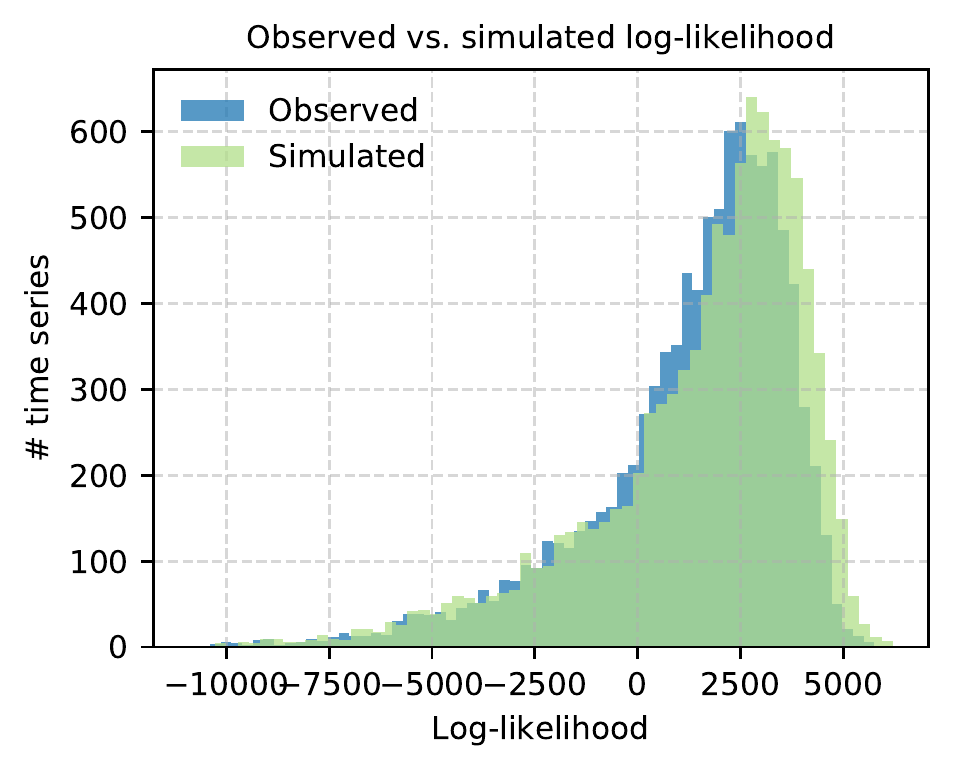}
    \caption{Distribution of observed and simulated log-likelihood on 100k time series.}
    \label{fig:log-likelihood-dist}
\end{figure}

We compare the distributions of the likelihood on observed and synthetic time series in Figures \ref{fig:log-likelihood-qq} (Q-Q plot) and \ref{fig:log-likelihood-dist} (histograms).
It can be seen that both distributions are similar, with the simulated time series being slightly more likely.
This demonstrates that the HDP-HMM explains well the diversity of observed trajectories in RIPE Atlas measurements. 

Thus, we have not only visually verified on a large number of series that the segmentation obtained with the model is consistent with what a human expert would produce (Section \ref{sec:first-look}). But in addition, we have checked on a very large scale (about 100k randomly chosen series among the Atlas mesh measurements) that all these series are well modeled by the HDP-HMM. 

\section{Conclusion}
\label{sec:conclusion}

In this paper we have shown that the HDP-HMM model, a hidden Markov chain model with a potentially infinite number of states, is a very promising method for analyzing RTT time series over the Internet on long time scales (hours to weeks). 
We have recalled the principles of this model that produces an accurate segmentation of time series and identification of hidden states. Unlike black box approaches, the HDP-HMM provides some explainable parameters that can be used as input in different network management tasks such as the choice of routes, QoS prediction, or optimization of the measurement strategy.

Segmentation results are very close to what a human expert would provide. But the analysis method is fully automated with no human intervention, even in the initialization phase, and it is scalable. As proof, it has been implemented on an Internet-wide operational measurement infrastructure, RIPE Atlas, with a publicly available Web API.

We have shown that this method can accurately detect moments when abnormal events occur on the Internet. In the future we would like to automate this detection, and in particular to locate anomalies (infrastructure failures, etc...) in a precise way. This will require the use of other methods exploiting the diversity of the measured paths and tomographic approaches or using a preliminary timeseries filtering strategy. We will also work on a real-time processing of measured data to detect novelties in RTT series with HDP-HMMs in an almost instantaneous way, based on some recent sequential approaches to inference in HDP-HMMs.

\bibliographystyle{ACM-Reference-Format}
\bibliography{bibliography}

\end{document}